# Intelligent Scoliosis Screening and Diagnosis: A Survey

Zhenlin Zhang[1], Lixin Pu[1,2▽], Ang Li[3], Jun Zhang[2], Xianjie Li[4], Jipeng Fan[2]

1 University of Electronic Science and Technology of China, Sichuan Chengdu 611731

2 Chengdu Chengdian Jinpan Health Data Technology Limited Company, Sichuan Chengdu 611731

3 Hospital of Chengdu University of TCM, Sichuan Chengdu 611731

4 Zigong First People's Hospital, Sichuan Zigong 643099

▽ Corresponding author, E-mail: plx@uestc.edu.cn

**Abstract** Scoliosis is a three-dimensional spinal deformity, which may lead to abnormal morphologies, such as thoracic deformity, and pelvic tilt. Severe patients may suffer from nerve damage and urinary abnormalities. At present, the number of scoliosis patients in primary and secondary schools has exceeded five million in China, the incidence rate is about 3% to 5% which is growing every year. The research on scoliosis, therefore, has important clinical value. This paper systematically introduces computer-assisted scoliosis screening and diagnosis as well as analyzes the advantages and limitations of different algorithm models in the current issue field. Moreover, the paper also discusses the current development bottlenecks in this field and looks forward to future development trends.

**Keywords**：Machine learning; Scoliosis; deep learning; Multimodal images

## 0 Introduction

Scoliosis[1], also known as spinal curvature or spinal deformity, is a three-dimensional abnormality of the spine [1], characterized by a Cobb angle [2] greater than 10°. Adam's forward bending test is commonly used for initial screening.[3] If a protrusion is observed on the back, scoliosis may be present, requiring X-rays to be taken and the Cobb angle to be calculated for further diagnosis. Typically, X-ray medical images of the human spine are taken in the sagittal and coronal planes, or full-length standing spinal X-ray images are used to assess the severity, nature, and progression of scoliosis, followed by quantitative evaluation.

The Cobb angle is a commonly used indicator for evaluating scoliosis. First defined by Cobb et al. in 1948,[2] it measures the maximum angle of a specific region of the spine from the upper endplate of the upper vertebra to the lower endplate of the lower vertebra. Since then, the Cobb angle has become the "gold standard" for assessing scoliosis. However, this method has some drawbacks:

(1) The Cobb measurement method only utilizes information from the two-dimensional plane of the spine, leading to measurement errors due to the lack of three-dimensional structural information.

(2) Doctors need to spend a significant amount of time and effort on line drawing and measurement operations on X-rays, and the results are highly subjective, depending on the experience of the clinician.

(3) Cobb angle measurements require X-rays of the patient as the basis for measurement, which can cause various types of harm to the patient due to X-ray exposure.[4, 5]

Despite these drawbacks, the method has clear advantages, such as standardized measurement methods and stable results with good diagnostic value. This paper provides a retrospective review of this field for the first time, categorizing solutions into those based on image processing, point cloud processing, and other methods. It analyzes the advantages and limitations of different approaches, explores the

1 Funding Projects: Key R&D Project of Regional Innovation Cooperation of Sichuan Provincial Department of Science and Technology: "Intelligent Diagnosis System for Scoliosis Based on Artificial Intelligence" (Project Number：2022YFQ0097）; Basic Research Project of Shenzhen Science and Technology Plan: "Key Technologies Research of AI-based Breast Cancer Intelligent Detection System" (Project Number：JCYJ20210324134405016)

▽：Corresponding Author，Email：plx@uestc.edu.cn

bottlenecks restricting the development of this field, and discusses future trends.

# 1 Spine Segmentation and Recognition

Spine recognition and segmentation are important topics in medical image processing, with significant implications for the prevention and screening of various diseases, including scoliosis. They also present a major challenge in the field of image segmentation. The introduction of spine segmentation tasks in the International Conference on Medical Image Computing and Computer Assisted Intervention (MICCAI) in 2019 and 2020 highlighted the significance of this area. Sekuboyina et al.[6] summarized the results of these competitions, where participants trained algorithm models using official CT datasets provided by the organizers. The challenge consisted of three stages: (1) Algorithm model training, (2) White-box testing, and (3) Black-box testing. In stages (1) and (2), participants proposed algorithms and measured their performance. Stage (3) involved black-box testing, where the performance of algorithms was evaluated based on closed testing by the organizers using submitted algorithm models. Such testing procedures place high demands on the generalization capabilities of algorithms.

The paper provides a brief introduction to 25 algorithms participating in the competition and summarizes four evaluation metrics: recognition rate, localization distance, Dice coefficient[7], and Hausdorff distance [8]. The first two metrics reflect the performance of the algorithms in localization and labeling, while the latter two metrics assess the performance in image segmentation.

Payer et al. [9] divided the overall task into three stages: (1) spine localization; (2) vertebrae localization and identification; (3) vertebrae segmentation. In stage (1), the authors employed a variant of U-Net[10] o regress the heatmap of the spine's centerline. In stage (2), they used the Spatial-Configuration-Net[11] to localize the center of each vertebra. In stage (3), U-Net was used for binary segmentation, achieving promising results. Their black-box test yielded a recognition rate of 94.25%, a localization distance of 4.8 mm, a Dice coefficient of 89.8%, and a Hausdorff distance of 7.08 mm, ranking first among all algorithms in 2019. Lessmann et al. [12] also utilized a U-Net network, with the main idea focused on iteratively shifting the Region of Interest (ROI) to perform image segmentation. This approach yielded good results, securing second place in the rankings.

n the 2020 MICCAI Spine Segmentation Challenge, Chen et al.[13] proposed a multi-stage solution based on a deep reasoning network. Initially, a simple low-resolution U-Net was used to determine the approximate region, followed by a high-resolution U-Net to perform binary segmentation of the entire spine. For vertebra segmentation, inspired by Lessmann et al.[12], the authors adopted an iterative 3D U-Net to segment the vertebrae from the background. They innovatively employed a deep reasoning framework to encode and constrain the model, effectively leveraging anatomical structures and prior information. Ultimately, their approach stood out among all algorithms. In the black-box test, the method achieved a recognition rate of 96.58%, a localization distance of 1.38 mm, a Dice coefficient of 91.23%, and a Hausdorff distance of 7.15 mm.

Chenhen Zhang[14] proposed an improved vertebrae segmentation model, DAU-Net, based on the U-Net architecture, which effectively enhanced segmentation efficiency and reduced the parameter size. The paper also introduced a semi-supervised framework for vertebrae segmentation, aimed at reducing the cost of manual annotation and addressing the issue of insufficient high-quality datasets.

Nicolaes [15] also employed a 3D convolutional network for segmentation, achieving an AUC of 93%. Tao et al.[16] localized the spinal curve by estimating landmark points, obtaining a favorable SMAPE score. There are numerous other studies in this area. However, the above methods rely on X-ray or CT images for spine localization and identification, which require additional auxiliary work when screening and diagnosing scoliosis, thereby increasing the task's workload.

# 2 Image Processing-Based Scoliosis Screening and Diagnosis

Image processing is currently the mainstream approach for intelligent scoliosis diagnosis and is a well-developed research direction. Most methods rely on convolutional neural networks (CNNs) to construct architectures, leading to the development of various other variants. In addition to directly using X-ray images, methods have been developed that utilize multimodal imaging and image generation techniques for diagnosis.

## 2.1 X-ray Image-Based Screening and Diagnosis

BoostNet [17] and MVC-Net [18] are two representative algorithm models used for scoliosis diagnosis based on X-ray images. The BoostNet model performs spinal landmark detection and localization on X-ray images by combining convolutional neural networks with statistical methods, using the detected landmarks for evaluation. On the other hand, the MVC-Net model divides the task into three stages, with different methods designed for each stage. It combines convolutional layers, a spinal key point estimator, and a Cobb angle estimator to output both spinal key points and Cobb angle measurements.

Fan Liu[19] summarized CNN-based scoliosis screening methods, starting by comparing the performance of models such as R-CNN, Fast R-CNN, and Faster R-CNN in target localization, specifically focusing on the localization of the patient's back. The study then compared the performance of various classic convolutional neural networks, including AlexNet, VGG, DenseNet, and ResNet. Based on the Cobb angle, scoliosis was classified into four categories: no scoliosis (Cobb angle 0–10°), mild scoliosis (Cobb angle 11–25°), moderate scoliosis (Cobb angle 26–45°), and severe scoliosis (Cobb angle >45°). The author used machine learning models for classification and ultimately selected a combination of Faster R-CNN and ResNet50 models, forming three binary classifiers. The achieved accuracy rates were 91.23%, 86.92%, and 82.45%, respectively.

Yongcheng Tu[20] proposed a traditional machine learning algorithm for the automatic measurement of Cobb angles from X-ray images. First, the enhanced watershed segmentation algorithm was used to segment the spine and extract the center points of each vertebra. Then, a sixth-order polynomial was employed to fit a curve to the set of center points, generating the spinal curve. Finally, the Cobb angle was calculated by determining the angles between the tangents at points where the second derivative of the spinal curve equals zero. Experimental results showed that the algorithm achieved a detection accuracy of 99.0%, a segmentation accuracy of 80.33%, and a Cobb angle measurement error of ±4.99°. This study also demonstrated that traditional machine learning algorithms can achieve promising results.

Caesarendra et al.[21] proposed a method for scoliosis diagnosis using CNNs, which utilizes input X-ray images to obtain the locations of seventeen vertebrae and subsequently processes this information to output the Cobb angle. This method achieved an accuracy of 93.6%, and its reliability closely matched that of clinical diagnoses.

X-ray image-based scoliosis diagnosis has long been regarded as one of the most reliable methods, and research in this area has become quite saturated, achieving high accuracy rates. However, the unavoidable exposure to radiation from X-rays poses risks to the human body, which is a challenge that other methods aim to address.

## 2.2 Screening and Diagnosis Based on Other Image Modalities

Utilizing 3D ultrasound images to estimate the X-ray Cobb angle (XCA) through the Spinous Process Angle (SPA) has garnered attention. Yang et al. [22] proposed a semi-automated method for analysing and measuring the spinal curvature angle based on the transverse process landmarks. This method demonstrated higher consistency compared to XCA, indicating that the "gold standard" for scoliosis diagnosis is no longer singular.

Ungi et al.[23] proposed a method for segmenting and visualizing tracked ultrasound images to measure spinal curvature. This method, based on convolutional neural networks, aims to overcome the limitations of tracking ultrasound. The authors discussed whether this automated segmentation approach could serve as a substitute for X-ray images in scoliosis diagnosis. However, a limitation of this study is the small sample size, as the network was trained using data from only eight volunteers and tested on another eight, lacking statistical significance.

The direction of multimodal image diagnosis requires further research. On one hand, there is a need to improve the accuracy and reliability of these methods; on the other hand, more exploration into other modalities of image processing is warranted.

### 2.3 Image Generation-Based Methods

In addition to directly using non-X-ray images, X-ray images can also be generated by incorporating additional information for screening and diagnosis. Wong et al.[24] proposed a method based on RGBD images, using RGBD images of the patient's back to generate X-ray images, which were then utilized for screening and diagnosis. The authors employed the Microsoft Azure Kinect DK depth camera to capture RGBD images of the patient's bare back in an upright position, marking six key points that provided important information about the spinal curvature, including the position of the sacrum. They used a High-Resolution Net[25] for learning and prediction. The High-Resolution Net is a common architecture for semantic segmentation and human pose estimation, which integrates CNNs of different resolutions in parallel to achieve multi-scale information fusion, resulting in more accurate predictions. The obtained six landmark points and the RGBD images were then input into a CGAN[26, 27] model. It is noteworthy that this study primarily focused on image generation rather than directly utilizing the generated images for diagnosis. Additionally, the authors employed data augmentation techniques to expand the dataset, including random translations, rotations, and scalings of the original dataset. This method effectively addresses issues related to overfitting and is a valuable approach for mitigating the problem of insufficient data.

Jiang et al.[28] proposed a method for generating X-ray images using images obtained from a 3D ultrasound imaging system. They introduced the UXGAN model based on the CycleGAN framework[29], incorporating attention mechanisms and residual blocks into the network architecture. The attention mechanism[30], initially proposed in the field of natural language processing, has also demonstrated its advantages in computer vision and related fields. Residual connections can effectively reduce model overfitting and accelerate the training process.

The image generation method employs generative adversarial networks (GANs), offering flexibility by allowing the generation of X-ray images from various types of input images. However, this approach can be complex and requires a substantial amount of high-quality datasets to train the model effectively.

## 3 Point Cloud Processing-Based Methods

Point cloud processing is a relatively novel approach that effectively addresses the issue of patient exposure to radiation during X-ray imaging, offering a simpler and faster diagnostic process.

Sudo et al.[31] were the first to utilize point cloud data for scoliosis screening. They employed a 3D sensor to capture the patient's flexed back and generate point cloud data of the back. Using algorithms such as 3D moving average filtering[32] and Iterative Closest Point (ICP) point cloud registration[33], they calculated the "asymmetry index" of the back point cloud. Finally, they predicted the Cobb angle by identifying the linear relationship between the asymmetry index and the Cobb angle. Experimental results indicated that the system required only 1.5 seconds for scanning and analysis, achieving an AUROC value of 0.96 for predicting Cobb angles greater than 25°. However, the study included only 76 patients for validation, indicating issues with sample size.

Kokabu et al.[34] further investigated the statistical relationship between the asymmetry index and the Cobb angle, expanding their experimental sample to 170 cases. The results showed a correlation coefficient of 0.88 between the asymmetry index and the Cobb angle. The study classified the Cobb angle into three categories: 15°, 20°, and 25°, conducting binary classification predictions for each category and achieving corresponding AUROC values of 0.92, 0.94, and 0.96, respectively. The paper also provided an in-depth analysis of the advantages and limitations of Sudo's[31] method, evaluating the impact of smoothing and registration operations on diagnostic outcomes. It emphasized that this algorithm does not rely on the patient's position or orientation, making it relatively stable, but it also noted a high requirement for the accuracy of the 3D camera used.

Kokabu et al.[35] , building on previous research, employed a deep learning approach by creating a CSV file of the differences between the left and right sides of the human back, which was then directly input into a convolutional neural network to predict the maximum Cobb angle. The correlation coefficient improved to 0.91, with an average prediction accuracy of 94% for the Cobb angle. However, compared to traditional methods, the overall improvement in performance was not substantial.

In summary, research in the area of point cloud processing is still limited, and the diagnostic accuracy requires further enhancement. Additionally, the diagnostic process itself needs to be standardized further.

# 4   Surface Topography System

The surface topography (ST) system is a method for assessing spinal curvature based on the external body contour, which can be implemented using various techniques. A classic example is the moiré fringe ST system[36], which processes the interference patterns projected onto the patient's back. Currently, most methods rely on computational image capture and digital analysis.

Using the ST system as a substitute for traditional radiographic methods in scoliosis screening[36-38] differs from the previously mentioned algorithms. The ST system is capable of analyzing minute deviations and changes in flat surfaces to obtain three-dimensional structural information. Chowanska[38] explored the potential of the ST system as a replacement for scoliosis measurement tools and concluded that this method lacks feasibility. However, Applebaum et al. [39] compared the advantages and limitations of the ST system with traditional radiographic techniques from historical, developmental, and effectiveness perspectives, concluding that the ST method could be improved and gradually replace traditional radiography for various disease diagnoses. The article notes that the ST system may yield different results for individuals of varying body types and may not provide the precise information that X-rays can. However, these limitations could be progressively addressed through techniques like raster stereography.

# 5   Summary and Outlook

## 5.1   Challenges and Limitations

Insufficient datasets are a pervasive problem in medical image processing. Several factors limit dataset size, including a small number of patients, challenges in obtaining labeled data, and the confidentiality of medical information. Additionally, the lack of generalizability of datasets is another significant issue in this field. Competitions and medical institutions are primary channels for collecting medical datasets, but the information contained in data gathered from different sources can vary significantly. This discrepancy arises from differences in data distribution, leading to variations in dataset performance and making it difficult to achieve generalizability.

Moreover, aside from using X-ray images for diagnosis, the accuracy and generalizability of other methods are often questionable. This issue is also rooted in the variability in data collection, as different diagnostic methods may have specific requirements for patients. Addressing these challenges is crucial for improving the reliability and applicability of medical imaging techniques in clinical practice.

## 5.2 Future Development

In summary, due to the varying sensitivities of different models and algorithms, as well as the differences between datasets, utilizing multimodal data for scoliosis assessment is likely to be a major focus of future research. In the coming years, more information may be leveraged for diagnosing scoliosis, and establishing a standardized database will be one of the primary challenges to address in the field of scoliosis diagnosis.

Additionally, semi-supervised learning represents a valuable research direction. By employing semi-supervised learning techniques, it is possible to achieve good predictive performance with minimal labeled datasets. However, current research in this area is still limited and requires further exploration.